\documentclass[11pt,twoside]{article}

\usepackage{asp2004}
\usepackage{epsf}
\usepackage{psfig}
\usepackage{lscape}

\begin{document}
\title{The Boundaries of the GW Vir Stars in the Effective Temperature
  - Surface Gravity Domain}   
\author{P.-O. Quirion$^{1,2}$, G. Fontaine$^1$, and P. Brassard$^1$ }   
\affil{$^1$ D\'epartement de Physique, Universit\'e de Montr\'eal, Montr\'eal, Qu\'ebec, Canada H3C 3J7 \\ $^2$ Institut for Fysik og Astronomi, Aarhus Universitet, {\AA}rhus~C, Danmark, DK-8000}
\begin{abstract} 
We derive the theoretical red edge of the pulsating GW Vir stars by using
full evolutionary calculations that involve mass loss and diffusion. 
We put the emphasis on the fact that the specific mass loss law used in
the evolutionary computations determines the red edge's position. By
combining this specific property with the observed location of the red
edge in the effective temperature-surface gravity domain, we obtain
interesting constraints on possible mass loss laws for PG 1159 stars.
\end{abstract}
\section{Astrophysical Context}
The GW~Vir stars constitute a class of pulsating stars descending from
post-Asymptotic Giant Branch (post-AGB) evolution. They exhibit
low-amplitude multiperiodic luminosity variations caused by low degree,
intermediate-order nonradial $g$-mode oscillations. Their observed
periods are found in the range from 300 s to upward of 5000 s. Although
there has been some debate in the past as to the exact cause of the
pulsational instabilities, it is now fair to say that there can no
longer be any doubt that the physical process responsible for the
excitation of $g$-modes in these stars is a $\kappa$-mechanism
associated with the partial ionization of the K-shell electrons in
carbon and oxygen, two important constituents of the envelopes of these
stars (see, e.g., Saio 1996, Gautschy 1997, Quirion, Fontaine, \&
Brassard \citeyear{QFB04}, and Gautschy, Althaus, \& Saio 2005). The
GW~Vir pulsators are found in a very wide area of the $\log g-T_{\rm eff}$
diagram covering the ranges $5.5 \la \log g \la 7.5$ and $80,000\ \rm{K}
\la T_{\rm eff} \la 170,000\ \rm{K}$. This area overlaps with two
spectroscopic classes: the Wolf-Rayet Central Stars of Planetary Nebula
([WCE]) and the hottest white dwarfs, the PG~1159 stars. As discussed in
the excellent review paper presented recently by Werner \& Herwig
(2006), quantitative spectroscopy and detailed evolutionary calculations
imply very strongly a tight evolutionary connection following the
sequence [WCE]$\rightarrow$PG~1159. In this context, it has been suggested that
pulsating stars from these two spectral types and sub-types should all be
included under the unique vocable of GW~Vir stars \citep{QFB06}.     

The one factor that differentiates the PG~1159 from the [WC] stars
despite their similar atmospheric compositions is the presence of much
stronger winds in the latter. The wind produces wide carbon emission
lines, giving the [WC] type its characteristic spectral signature. With
time and decreasing luminosity, the magnitude of the outgoing wind
diminishes, thus weakening the carbon emission lines in the [WC] stars
and letting the PG~1159 features appear in the spectra of the [WC]-PG~1159 
transition objects. A further decrease of the mass loss and wind
completes this scenario whereby the wide carbon emission lines
disappear, leading to a full characteristic PG~1159 spectrum.  

The chemical composition of the envelope of GW~Vir stars is critical for
the determination of their stability. It has been shown in \cite{QFB06}
that the temperature of the blue edge drawn for this class is directly
dependent on the quantity of carbon and oxygen present in their
envelope. In general, the blue edge moves from lower to higher
temperatures as the quantity of carbon and oxygen increases. When
considering the dispersion in atmospheric parameters and in chemical
composition in GW~Vir stars, we must understand that the blue edge is a
''fuzzy'' notion for this type of pulsating stars. The positions of the
blue edges for GW Vir stars, calculated for a large spectrum of chemical
composition, can be found in \citet{QFB06}.    

The evolutionary connection following the sequence PG~1159$\rightarrow$DO, 
where the spectral type DO belongs to He-dominated atmosphere white
dwarfs cooler than $\sim$ 80,000 K, is also strongly supported by
quantitative spectroscopy \citep{WerHer06}. The strong wind present in
[WC] and PG~1159 stars tends to homogenize their envelope \citep[see,
  e.g.,][]{Cha97,UngBue01}, but as this wind weakens with lowering
luminosity, it progressively looses its homogenizing capacity and
gravitational settling takes over and will cause carbon and 
oxygen to precipitate while helium will float at the surface of the
star. As carbon and oxygen are responsible for the $\kappa$-mechanism in
these stars, it is expected that gravitational settling will ultimately
stop pulsational driving in GW~Vir stars, drawing the red edge of this
class. We discuss here representative calculations showing the effects
of the interaction between diffusion and mass loss on the position of
the red edge of GW~Vir stars.       

\section{Results and Discussion}

We used a improved version of the evolutionary code based on a finite
element method introduced in \citet{Fon01} to model the effects of
diffusion and mass loss on the red edge's position. Diffusion is treated
by resolving the following mass conservation equation  
\begin{equation}\label{eq:diff}
\frac{\partial n_i}{\partial t}= -\frac{1}{r^2} \frac{\partial}{\partial
  r} r^2 \left[n_i(w_i+W)\right],
\end{equation} 
where $n_i$, $t$, $r$, $w_i$, $W$ are respectively, the number abundance of
element $i$ ($=$ He, C, O in the case of GW~Vir stars), the time, the
radial coordinate, the velocity of element $i$, and the wind velocity. The
microscopic velocity $w_i$ of each element is calculated from first
principles (effects of gravity, composition gradient, temperature
gradient, and electric field) as explained in \citet[][]{IbeMac85}.  

On the other hand, we cannot compute the macroscopic wind speed $W$ from
first principles, as there is no complete physical theory addressing this
problem yet. The usual method to model a stellar wind is to scale the mass
loss of a star on its luminosity. When a suitable mass loss law has
been selected, we link it to the wind speed $W$ via flux conservation  
\begin{equation}\label{eq:mloss}
W=\frac{\dot{M}}{4\pi r^2\rho},
\end{equation}
where  $\dot{M}$ is the mass loss rate and $\rho$ the density. In this
equation, the mass loss is assumed to be small enough to have negligible
effects on the total mass of the star. 

The mass loss laws used in the present calculation are given in
Table~\ref{tab:mloss}. WM1 is a fit to the mass loss rates measured in five PG
1159 type stars --- three of which are also GW~Vir stars --- published in
\citet{KoeWer98}. WM2 is a fit to the five previous stars plus nine
nuclei of planetary nebulae of similar luminosity found in
\citet{Pau04}. For WM3, we chose the theoretical model used for
post-AGB calculations in \citet{Blocker95}. Finally, WM4 is an empirical
law derived in such a way that the theoretical red edge falls directly on the
empirical red edge observed for the GW~Vir class.   

\begin{table}[!ht]
\begin{center}
\caption{ The Four Wind Models \label{tab:mloss}}
\smallskip
\begin{tabular}{cc}
\tableline
\tableline
\noalign{\smallskip}
{Model} & $\dot{M}\ [M_{\rm \odot}$ yr$^{-1}]$\\  
\tableline
\noalign{\smallskip}
WM1 & $1.14\times 10^{-11} L^{0.93}$\\
WM2 & $1.82\times 10^{-13} L^{1.36}$\\
WM3 & $1.29\times 10^{-15} L^{1.86}$\\
WM4 & $1.00\times 10^{-17} L^{2.38}$\\
\tableline
\end{tabular}
\end{center}
\end{table}

\begin{figure}[!ht]
\plotfiddle{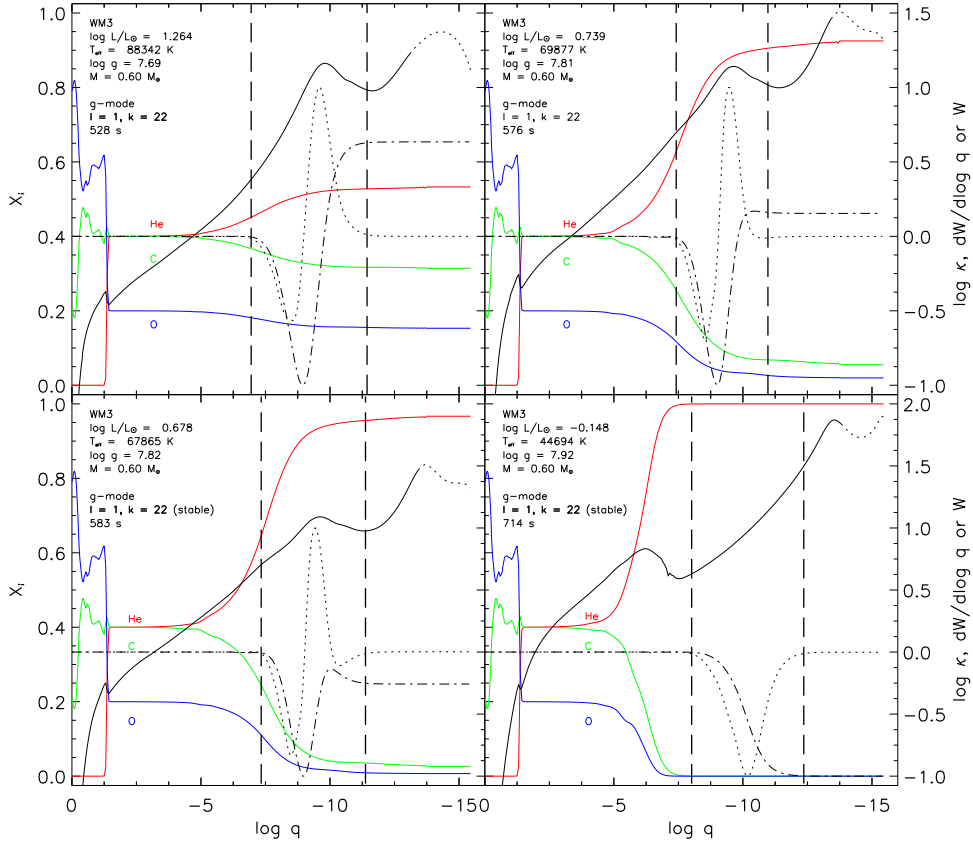}{11cm}{90}{65}{65}{255}{-10}
\caption{ Four models taken along a $0.6~ M_{\rm \odot}$ track showing the
  effects of diffusion on the chemical composition of the envelope of a
  star undergoing mass loss according to the WM3 law. The composition of the
  envelope at the start of the calculation was $X$(He) = 0.40, $X$(C) =
  0.40 and $X$(O) = 0.20 in mass units.  The solid curves associated with
  helium, carbon and oxygen (in red, green and blue respectively in the
  electronic version) depict the progressive rise of the helium content
  at the surface, prompted by gravitational settling of carbon and oxygen,
  as the mass loss decreases with decreasing luminosity. The two models
  in the top panel, at $T_{\rm eff} = 88,342$ K and $69,877$ K, are unstable,
  as shown by the positive value of the work integral $W$ (running from
  left to right) at the surface of the models. The driving/damping
  region, defined by the zone between the two vertical dashed lines, indicates
  where the magnitude of the derivative of the work integral $dW/d\log q$ is
  greater than 1\% of its maximal value (normalized at either $+ 1$ or $-
  1$). The maximum of the driving occurs, as is always the case for a classic
  $\kappa$-mechanism, near the maximum of  the opacity bump. The bump is
  seen in the Rosseland  opacity $\kappa$ around $\log q \sim - 9.5$. For the
  two stable models in the lower panels, at $T_{\rm eff} =  67,865$ K and
  $44,694$ K, we see negative values of the  work integral at the surface
  of the star along with  the vanishing opacity bump. The dotted
  extensions of  the $\kappa$ curves mark the extent of the atmosphere,
 defined here as those layers with an optical depth  $\tau_{Ross}<100$. 
\label{fig:quatre}}
\end{figure}

In Figure~\ref{fig:quatre} we chose four temperatures along our 0.6
$M_{\rm \odot}$ evolutionary track, under the effect of diffusion and WM3, to
represent, in a chronological way, the evolution of the internal
structure of our model and the formation of the red edge. We see that,
as the star cools down, the carbon and oxygen abundance in its
driving/damping layers get lower. This reduction leads to a smoothing of
the opacity bump created by the partial ionization of C/O, making the
$\kappa$-mechanism, responsible for the pulsations, less and less
efficient. The red edge is drawn in between the 69,877~K and the
67,865~K models. It is interesting to note that the last unstable model,
at 69,877 K, shows only a small amount of carbon and essentially no
oxygen at its surface with $X$(He) $\simeq$ 0.97 and $X$(C) $\simeq$
0.03. This composition ratio would prevent a GW~Vir model with a
strictly uniform envelope composition (from the surface to below the
driving/damping region) to be unstable. We come back on this phenomenon in the
following paragraphs. The coolest model, stable at 44,694~K, has been
transformed into a DO white dwarf. Indeed, helium completely fills the
envelope from the surface down to $\log q\sim-7.0$, below the driving
region. The residual opacity bump at $\log q\sim -6.0$ only marks the
transition from almost pure helium to mixed He/C/O in deeper layers.      

The qualitative picture of gravitational settling with the WM3 model is
representative of what is happening with the other wind models. With the
more important mass loss rates of WM1 and WM2, however, the settling of
C and O is comparatively slowed down and the red edge is pushed to lower
temperatures. By the same token, a hotter red edge is found for the
weaker mass loss rate associated with the WM4 law. We mention
that for models initially unstable above 100,000 K, the
position of the red edge has only a marginal dependence on the initial
chemical composition in the envelope of the model. Thus, the dominant factor
affecting the position of the red edge is the mass loss law. A corollary to
this affirmation is that the average magnitude of the mass loss in GW~Vir stars
(and more generally in PG~1159) could be estimated from the actual
position of the red edge.

\begin{figure}[!ht]
\plotfiddle{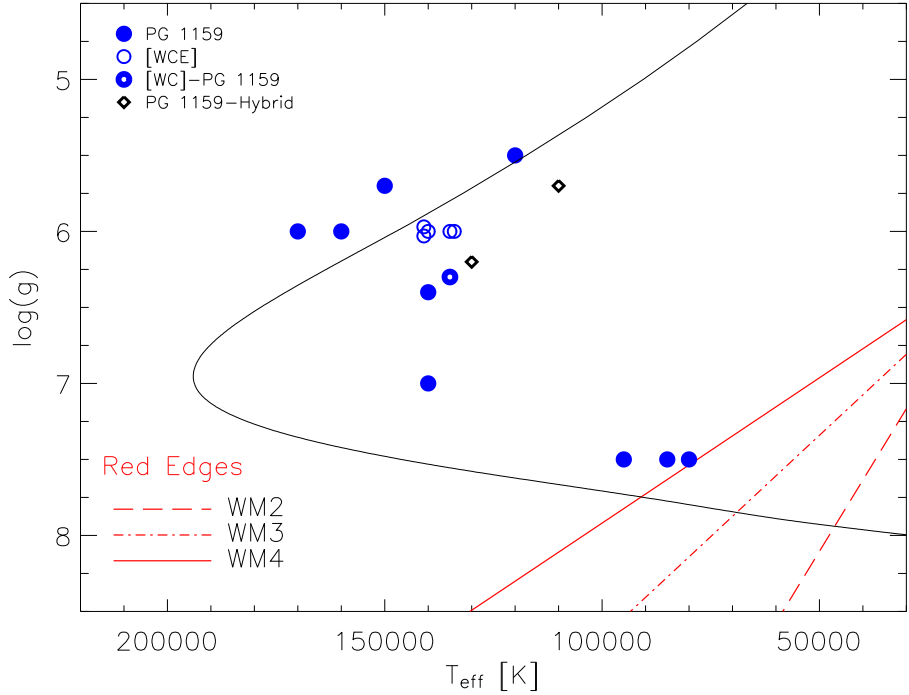}{9.3cm}{90}{65}{65}{210}{-10}
\caption{ Positions of the known GW~Vir stars in the $\log g-T_{\rm eff}$
  diagram. The four spectroscopic types and sub-types of GW~Vir are shown
  along with a representative 0.604~$M_{\rm \odot}$ track from F. Herwig and
  the calculated red edge for WM2, WM3 and WM4.    
\label{fig:hr}}
\end{figure}

The effects of the mass loss over the position of the red edge is
pictured in Figure~\ref{fig:hr}. We have calculated the evolution of
three different models having masses of 0.5, 0.55 and 0.6~$M_{\rm \odot}$. These
models were all allowed to evolve under the effects of the WM1, WM2, WM3, and
WM4 wind models. We then used our nonadiabatic pulsation code, briefly presented in
\citet{Fon94}, to probe the stability of the models along each track. The red 
edges obtained in the figure are simply fits along the three tracks
calculated with the different mass loss laws WM2, WM3 and WM4. The
position of the WM1 red edge, around $T_{\rm eff}=30,000$~K, is not shown here
as it is offscale. As is well known, no PG~1159 stars exists at this low
temperature. This temperature is clearly too cool and that particular
mass loss law must be abandoned. The same conclusion could be drawn for
WM2, but we prefer to set conservatively this law as a maximum value for
the magnitude of the mass loss in GW~Vir stars, $\dot{M} < $ WM2.  

By construction, and as indicated above, we have devised the WM4 model
in order to match fairly closely the empirical red edge as defined by
the position of the coolest known GW Vir star, PG~0122+200 at
80,000~K. However, it should be noted that the predicted surface
composition according to the WM4 model at that effective temperature is
highly deficient in carbon and oxygen as compared to the real
atmospheric chemical composition of PG~0122+200 ($X$(He) = 0.43, $X$(C)
= 0.39, and $X$(O) = 0.17). Hence, it would appear that the WM4 model
underestimates the true average mass loss in the GW Vir stars. On the
other end, the empirical red edge could also be actually somewhat cooler than the
effective temperature of PG~0122+200. In any case, we can use the WM4
model to set a minimum value for the mass loss in GW~Vir stars, $\dot{M} >
$ WM4. 

On the other hand, we see that the WM3 wind model depicted in Figure \ref{fig:quatre},
still permits relatively high abundances of carbon and oxygen in its 88,342 K unstable
model, situated close to the empirical red edge. Also, the red edge produced
by this wind model is not far, in the $\log g-T_{\rm eff}$ diagram, from the
coolest known GW~Vir stars. This model with $\dot{M}=$ WM3, is therefore
more likely to be representative of the actual red edge of GW~Vir stars.     
  
This brief analysis could be extended by taking into account the
position of the hottest DO white dwarfs. However, the cornerstone for the
accurate determination of the position of the red edge, and perhaps also
the precise determination of the actual mean mass loss rate in GW~Vir
stars, would be the discovery of a low carbon and low oxygen atmosphere PG~1159 or
DO star, showing luminosity variations caused by nonradial gravity modes.    
Such an object would have to reside almost exactly {\sl at} the red edge itself.


\end{document}